\documentclass{article}

\usepackage{microtype}
\usepackage{graphicx}
\usepackage{subcaption}
\usepackage{booktabs} %
\usepackage{amsmath}
\usepackage{marginalia}
\usepackage{stmaryrd}

\usepackage{hyperref}

\usepackage[nolist]{acronym}  %
\usepackage[capitalize]{cleveref}
\usepackage{enumitem}

\newcommand{\ATTN}[1]{\textcolor{blue}{#1}}
\newcommand{\todo}[1]{\textcolor{red}{#1}}
\newcommand{\TODO}[1]{\textcolor{red}{#1}}

\newcommand{\CUT}[1]{\textcolor{gray}{#1}}
\newcommand{\IGNORE}[1]{}

\renewcommand{\ATTN}[1]{#1}
\renewcommand{\todo}[1]{}
\renewcommand{\TODO}[1]{}

\renewcommand{\CUT}[1]{}

\newcommand*{\Opcodes}{x^o}

\newcommand*{\OpsEmbed}[0]{\mathbf{X}}

\acrodef{TPU}{Tensor Processing Unit}
\acrodef{DLA}{dense linear algebra}
\acrodef{XLA}{Accelerated Linear Algebra}
\acrodef{PCIe}{PCI Express}
\acrodef{HBM}{High Bandwidth Memory}
\acrodef{MAPE}{mean absolute percentage error}
\acrodef{HLO}{High-Level Optimizer}
\acrodef{LLO}{Low-Level Optimizer}
\acrodef{NAS}{Neural Architecture Search}
\acrodef{GAT}{Graph Attention Network}
\acrodef{GNN}{Graph Neural Network}

\usepackage[accepted]{mlsys2021}

\mlsystitlerunning{A Learned Performance Model for Tensor Processing Units}

\begin{document}

\twocolumn[
\mlsystitle{A Learned Performance Model for Tensor Processing Units}

\mlsyssetsymbol{equal}{*}

\begin{mlsysauthorlist}
\mlsysauthor{Samuel J. Kaufman}{gr,uw,equal}
\mlsysauthor{Phitchaya Mangpo Phothilimthana}{gr,equal}
\mlsysauthor{Yanqi Zhou}{gr}
\mlsysauthor{Charith Mendis}{gr}
\mlsysauthor{Sudip Roy}{gr}
\mlsysauthor{Amit Sabne}{gr}
\mlsysauthor{Mike Burrows}{gr}
\end{mlsysauthorlist}

\mlsysaffiliation{gr}{Google, Mountain View, CA}
\mlsysaffiliation{uw}{Paul G.~Allen School of Computer Science \& Engineering, University of Washington, Seattle, WA}

\mlsyscorrespondingauthor{Samuel J. Kaufman}{kaufmans@cs.washington.edu}
\mlsyscorrespondingauthor{Phitchaya Mangpo Phothilimthana}{mangpo@google.com}

\mlsyskeywords{Machine Learning, MLSys}

\vskip 0.3in

\begin{abstract}
Accurate hardware performance models are critical to efficient code generation.
They can be used by compilers to make heuristic decisions, by superoptimizers as a minimization objective, or by autotuners to find an optimal configuration for a specific program.
However, they are difficult to develop because contemporary processors are complex, and the recent proliferation of deep learning accelerators has increased the development burden.
We demonstrate a method of learning performance models from a corpus of tensor computation graph programs for \ac{TPU} instances.
We show that our learned model outperforms a heavily-optimized analytical performance model on two tasks---tile-size selection and operator fusion---and that it helps an autotuner discover faster programs in a setting where access to TPUs is limited or expensive.
\end{abstract}
]

\printAffiliationsAndNotice{\mlsysEqualContribution} %

\section{Introduction}

\begin{figure*}[t]
\begin{minipage}{.6\textwidth}
  \centering
  \includegraphics[width=1.\linewidth,trim={2cm 2cm 3cm 3cm},clip]{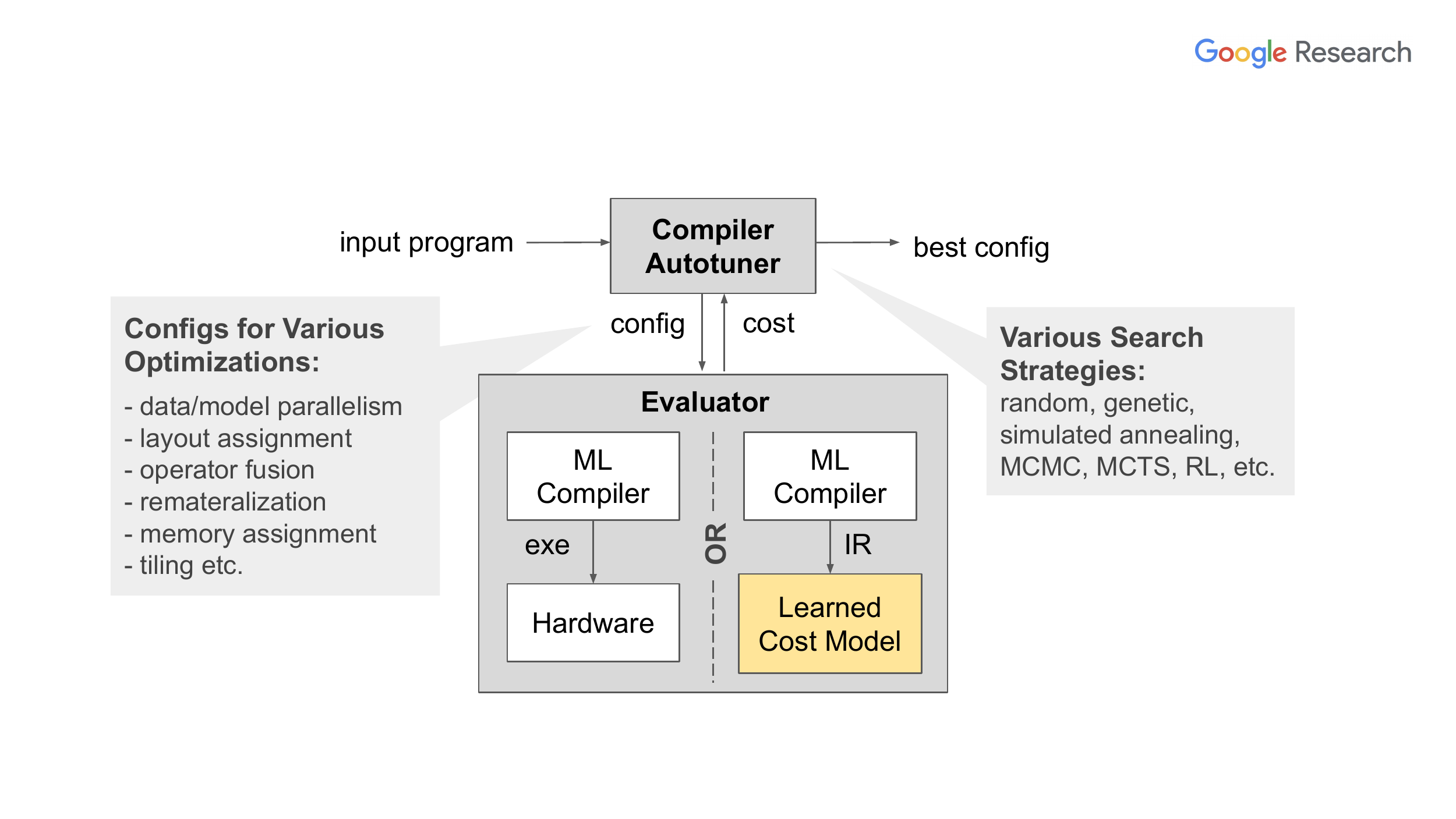}
    \caption{A compiler autotuner typically relies on real hardware to evaluate the performance of generated code. We propose a learned performance model as a cheaper alternative to obtain reward signals.}
    \vspace{-1em}
    \label{fig:autotuning-process}
\end{minipage}
\hfill
\hfill
\begin{minipage}{0.37\textwidth}
  \centering
  \includegraphics[width=1.\linewidth,trim={0cm 0cm 10cm 3cm},clip]{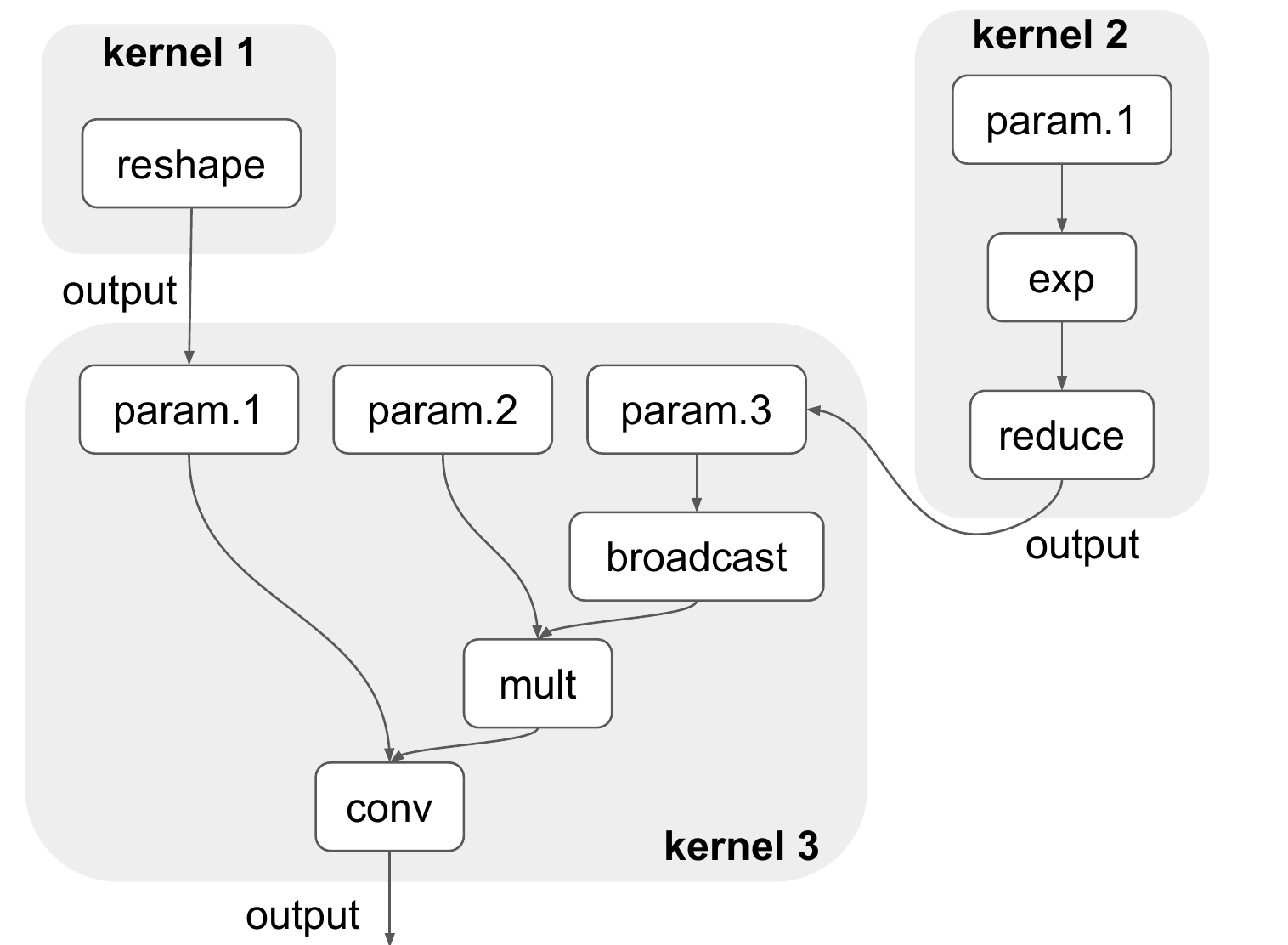}
  \caption{An optimized tensor computation graph consists of multiple kernels (gray blobs). Each kernel in turn contains a graph of nodes corresponding to primitive operations.}
    \vspace{-1em}
\label{fig:xla-graph}
\end{minipage}
\end{figure*}

Compilers often rely on performance models for solving optimization problems because collecting performance measurements from a real machine can be 
expensive, limited by hardware availability, or \CUT{even }infeasible (such as during ahead-of-time compilation).
For example, LLVM's loop vectorizer uses a performance model to compute the optimal vectorization and unroll factors~\cite{LLVM-vector}, and GCC uses a model to decide when to apply loop-peeling, loop-versioning, outer-loop vectorization, and intra-iteration vectorization~\cite{GCC-vector}.
In addition, a performance model can be used by a compiler autotuner to evaluate candidate configurations in a search space~\cite{AutoTVM,autohalide,PipeDream,roc}.

Developing an accurate analytical model of program performance on a modern processor is challenging and can take months of engineering effort. 
Program performance is tightly coupled with the underlying processor architecture as well as the optimization decisions that are made during compilation~\cite{Chaos}. Developers of analytical models are often unaware of detailed features of the processor or effects from all compiler passes. Furthermore, architectural features and the underlying compiler code generation interact in extremely complex ways; manually implementing these interactions and their effects on program performance is tedious and error-prone.
The recent proliferation of deep learning accelerators has only exacerbated this problem by demanding rapid, repeated development of performance models targeting new accelerators. 

This paper addresses these problems by applying machine learning techniques to produce a performance model. In particular, we are interested in learning a model for predicting execution time of tensor programs on TPUs, which are \ATTN{widely-used} accelerators for deep learning workloads~\cite{TPU-isca,TPU-training}. %
We aim to develop a learned approach to performance modeling that satisfies the following key criteria for the ease of development and deployment.
First, the approach must be general enough to handle non-trivial constructs in tensor programs (e.g., multi-level loop nests common in programs involving high-dimensional tensors).
Second, it must generalize across programs of different application domains as well as to programs unseen at training time.
Third, it should not rely on well-crafted features that require significant domain expertise and effort to develop and tune. Finally, the approach should be retargetable to different optimization tasks with minimal effort.

\IGNORE{
In this paper, we address these problems by applying machine learning techniques to produce a performance model for predicting execution time of tensor programs on a deep learning accelerator. Learned approaches to performance modeling need to satisfy four key criteria to make them practically viable for deployment in a real compiler or autotuner. First, the approach must be general enough to handle non-trivial constructs in tensor programs (e.g., multi-level loop nests common in programs involving high-dimensional tensors). Second, it must generalize across tensor programs of different application domains as well as to programs unseen at training time. Third, it \ATTN{must} not rely on well-crafted features that require significant domain expertise and effort to develop and tune. Finally, the approach should be retargetable to different optimization tasks with minimal effort.
}

While there has been some prior work~\cite{autohalide,AutoTVM,ithemal} proposing learned approaches to performance modeling, to the best of our knowledge, none of them satisfy the four criteria stated above. For instance, Ithemal~\cite{ithemal} does not handle complex multi-level loop nests. 
While Halide's learned performance model can handle tensor programs~\cite{autohalide}, it requires heavy feature engineering. Although AutoTVM's models do not rely entirely on manually-engineered features~\cite{AutoTVM}, it shows limited ability to generalize across kernels. 

Like prior work, we formulate the runtime estimation problem as a regression task. However, we make specific architectural choices to satisfy the desiderata.
First, our approach represents tensor programs
as data flow graphs with nodes that represent operations and edges that represent tensor flows between nodes. 
Second, we use a graph-based neural network \ATTN{optionally} coupled \ATTN{with a sequence model}; the graph model ensures generalizability across different programs, while the sequence model is used to capture long range dependencies within a graph. 
Third, we directly encode operation properties to generate a feature vector for a node in the graph.
While our approach does not require any program analyses, adding manually engineered features as additional features is trivial.
Our approach is retargetable to different tensor graph optimization tasks.
We evaluate our performance model on its ability to predict runtimes for two tasks: \emph{tile-size selection} and \emph{operator fusion}.
The model is applied to evaluate program configurations generated by an autotuner for the \ac{XLA} compiler~\cite{XLA} as depicted in~\cref{fig:autotuning-process}.

In summary, we make the following contributions:
\vspace{-1em}
\begin{itemize}
\itemsep0em 
    \item We develop a learned performance model for tensor programs that does not require feature engineering, generalizes to unseen programs, and is retargetable for different compiler optimization tasks.
    \item 
    We show that our learned models achieve 96.3\% and 95.5\% accuracy with respect to true measurements; and 2.4\% and 26.6\% better accuracy than the best hand-tuned model for tile-size and fusion tasks, respectively.
    \item We conduct a comprehensive set of ablation studies over modeling choices.
    \item We integrate our learned performance model into an \ac{XLA} autotuner, and demonstrate that it helps in discovering faster programs when access to real hardware is limited or expensive, which is often true in practice.
\end{itemize}

\section{Target Hardware and Tasks}

Our approach to learning a performance model is applicable to any target processor executing tensor programs.
A tensor program can be represented as a \emph{computation graph}, which is acyclic and directed. A node in a computation graph represents a tensor operation, processing one or more input tensors into a single output,
and an edge connects an output tensor from one node to an input tensor of another node.

To evaluate our method, we build a learned model to predict runtimes of XLA programs on a \ac{TPU}.
XLA is a machine learning compiler for multiple hardware targets, and is used by various machine learning programming frameworks. %
XLA first performs high-level optimizations at the whole-program level. During this stage, some nodes (primitive operations) in the original computation graph may be merged into a fused node, called a \emph{kernel}, as illustrated in \cref{fig:xla-graph}. After that, XLA lowers each kernel into a low-level representation, which is then further optimized and compiled to machine code.
In this paper, we
evaluate on two optimization tasks: tile-size selection (a kernel-level optimization applied during lowering) and operator fusion (a program-level optimization).

\subsection{Tensor Processing Unit}
\label{sec:accelerator}

Tensor Processing Units~\cite{TPU-training} are fast, energy-efficient machine learning accelerators. They achieve high performance by employing systolic array-based matrix multiplication units. 
The architecture incorporates a vector processing unit, a VLIW instruction set, 2D vector registers, and a transpose reduction permute unit.
Programs can access the \ac{HBM} or the faster but smaller on-chip scratchpad memory that is software-managed.
\ATTN{While a TPU has no out-of-order execution, it relies heavily on instruction-level parallelism---done by the compiler backend across several passes including critical path scheduling and register allocation---making it challenging for performance modeling.}
TPUs do not support multi-threading; one kernel is executed at a time.

The design of a \ac{TPU} allows us to compute the runtime of an entire program by summing the runtimes of its kernel executions.
\ATTN{We expect our approach to work best for targets where the runtime of one kernel is independent of others (e.g., no overlapping kernel executions and no inter-kernel caching).} For example, prior work has shown that this approach is sufficiently accurate for autotuning graph rewrites \cite{MetaFlow} and parallelization configurations~\cite{FlexFlow,PipeDream} on GPUs.

\ATTN{We evaluate our approach on TPUs v2 and v3 to demonstrate its generalizability across different generations of hardware. TPU v3 has higher memory bandwidth and twice as many matrix multiplier units compared to TPU v2.}

\subsection{Optimization Tasks}
\label{sec:tasks}

\paragraph{Tile-Size Selection}
\ATTN{To generate efficient code, XLA utilizes the fast scratchpad to store data.
Because of the limited scratchpad memory, a kernel cannot consume its whole input or compute its entire output at once. Instead, it computes one piece of its output at a time from one or more pieces of its inputs. These pieces are called \emph{tiles}. An output tile is copied to the slower HBM before the next tile is computed.} 
\ATTN{The goal of tile-size selection is to choose a tile size that minimizes kernel runtime.}

\IGNORE{Ultimately, we would like to replace this manual model with the learned performance model, demonstrating a new, cheaper way of developing compilers.}

\paragraph{Operator Fusion}
Operator fusion merges multiple operations into a single unit.
Before this pass, a node in a computation graph is a primitive tensor operation (e.g., convolution, element-wise add, etc.).
When producer and consumer nodes are fused, intermediate data is stored in scratchpad memory, without transferring it to/from HBM, thereby reducing data communication.
After the fusion pass, a node in a computation graph is either a single primitive operation or a fused operation with many primitive operations.

\subsection{Existing Analytical Model}
\label{sec:manual-cost}
For tile-size selection, XLA enumerates all possible tile sizes and selects the best according to a heavily hand-tuned analytical performance model. %
This model estimates the kernel's data transfer time and computation time, and takes the maximum of the two. Tile-size selection happens prior to the code generation, so the model relies on several heuristics that may cause inaccuracy. While the model works well in practice, the approach has the following demerits: (a) the lack of some execution behaviors due to poorly understood architectural characteristics implies missed opportunities; (b) changes in the code generation result in a constant need to update the heuristics; and (c) each new hardware generation requires not only tuning of existing heuristics but also additional modeling. Details can be found in \cref{appendix:manual-model}.

Unlike tile-size selection, XLA does not use a precise performance model for the fusion task. Instead, it relies on estimates of whether including each node into a fused group will save memory space and access time. It then prioritize fusion decisions according to these estimates.

\subsection{Autotuning}

\ATTN{
Instead of relying on the compiler's heuristics and analytical performance model, an XLA autotuner has been developed to search for the fastest tile size for each kernel, and the fastest fusion configuration of each XLA program.
The autotuner found up to 25\% speedup over the compiler's default on some production deep learning models. However, the autotuning process involves exhaustively running each kernel with all valid tile sizes (ranging from 2 to 500,000 options) and exploration in an exponentially large space of up to $2^{40,000}$ fusion configurations per each program. This requires many evaluations, where each evaluation is slow due to the time spent in compilation and execution.}
An accurate performance model can provide a cheap and reliable estimate of the runtime, and can significantly reduce the time and resource requirements of the autotuning process.

\section{Model Design}
\label{sec:model-design}
Our approach decomposes an XLA program into smaller computation subgraphs (kernels) whose runtimes are predicted with a neural network.
The estimated program runtime is the sum of its kernel runtimes.
Predicting kernel runtimes instead of the whole program's runtime has multiple benefits.
First, this decomposition is general enough that we can apply the neural network model to various tasks, including both program- and kernel-level optimizations.
Second, predicting runtime at a low-level representation should be more accurate as the model does not have to capture what happens inside the high-level compiler.
Additionally, kernels are smaller than whole programs, simplifying the model's domain.
The rest of this section focuses on the three main components for predicting a kernel runtime: model inputs, model architecture, and training objectives.

\begin{figure}[t]
  \centering
  \includegraphics[width=.9\linewidth,trim={0cm .42cm 12cm 1.2cm},clip]{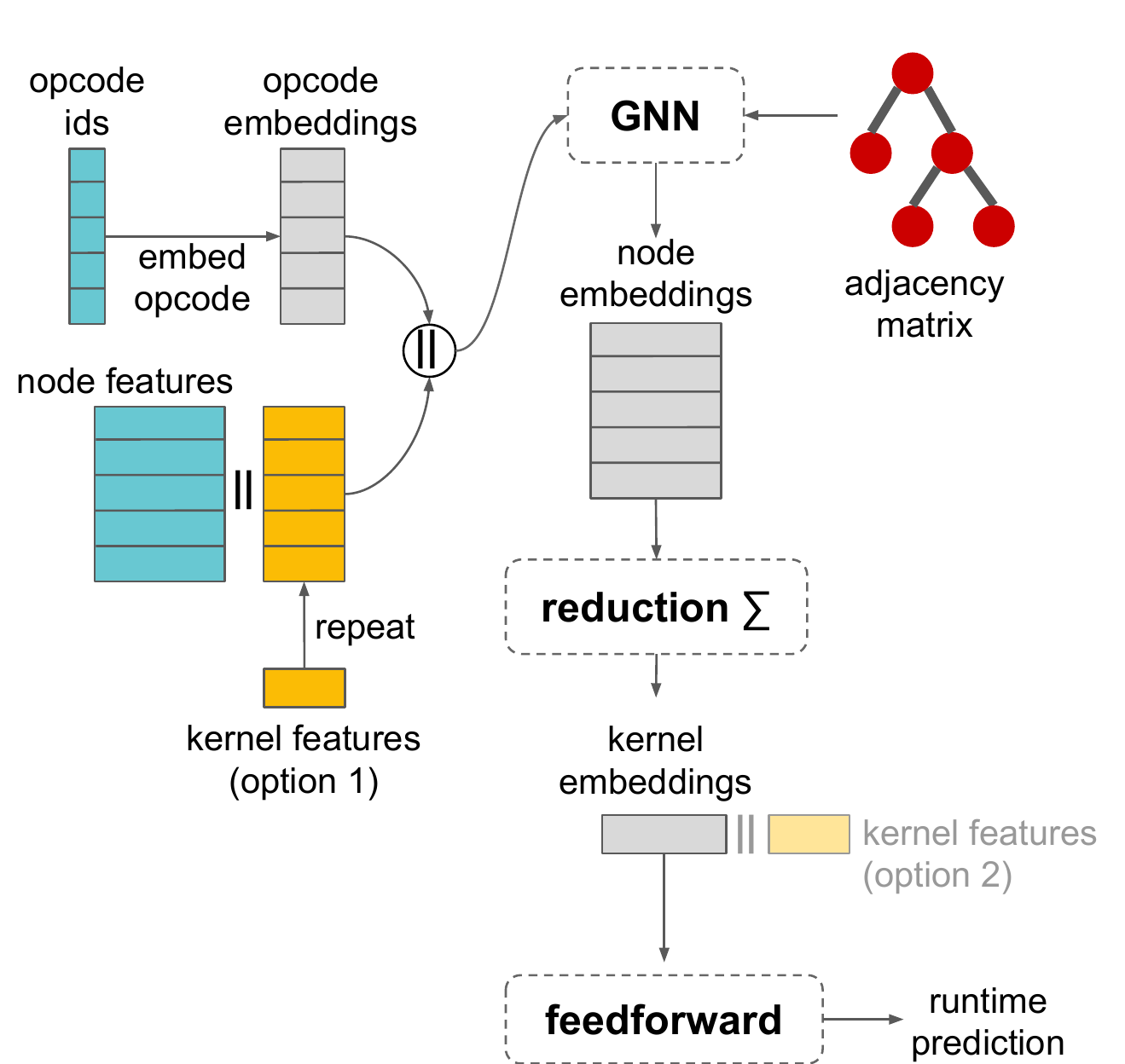}
  \vspace{-.5em}
  \caption{Overview of our model. Model inputs are an adjacency matrix representing graph structure (red), node information (blue), and kernel information (yellow). Kernel features can be fed into the model in two ways: (1) appending to the node features or (2) appending to kernel embedding.}
  \vspace{-.5em}
\label{fig:overview}
\end{figure}

\subsection{Model Inputs}
\label{sec:model-inputs}

A model input is a kernel represented as node features, whole-kernel features, and an adjacency matrix (highlighted yellow, red, and blue respectively in \cref{fig:overview}).
Node features include the integer-valued type of the operation (opcode), as well as scalar features which further describe the node's behavior, such as output tensor shape, tensor layout, striding, padding, and when applicable, convolution filter size.\footnote{\ATTN{Integers are cast to reals. Features are independently scaled to be in the range $[0, 1]$ using the minimum and maximum observed in the training set.
}}
Kernel's inputs are expressed by nodes with the \emph{parameter} opcode, and outputs are expressed via an extra feature associated with the output nodes.
Kernel features include tile size (only for the tile-size selection task) and optional static performance information.
The adjacency matrix captures data-flow dependencies between nodes in the kernel, as shown in \cref{fig:xla-graph}.

\paragraph{Optional Static Performance Features}
The XLA compiler has static analyses that determine high-level performance metrics of a given kernel.
In addition to the features extracted directly from the program representation, we consider providing information from these analyses as additional inputs to the model. These features are optional, but may improve the model's accuracy.
We consider four such kernel features: (1) number of floating point operations, (2) amount of data read in bytes, (3) amount of data being written in bytes, and (4) number of instructions executing on a special functional unit. These are estimates because the static analyses do not precisely model the compiler's backend code generation.
The static performance features of the same kernel with different tile-sizes are the same.

\paragraph{Variable-Sized Features}
Many node and kernel features are naturally interpreted as variable-length lists of numbers.
This is because tensors are $n$-dimensional arrays and some features describe individual tensor dimensions.
For example, tile size is encoded as a vector of length $n$, in which each component corresponds to a tensor dimension.
We encode these features as fixed-size sub-vectors, padded or truncated as necessary.
Additionally, we include the sum and product of all the values.
Including the product is critical as it usually represents the volume of a tensor, and can be more predictive in cases where the sub-feature has been truncated and so the product could not be recovered by the model.

\subsection{Model Architecture}
\label{sec:model-architecture}

\Cref{fig:overview} depicts the architecture of our model. We apply a \ac{GNN} to capture local
structural information and then apply a reduction over node embeddings to generate a kernel embedding, which is in turn used to predict the final runtime. We explore different choices for the reduction, including sequence and attention models that can capture global graph information.

\paragraph{Node and Kernel Features}
The opcode $\Opcodes_i$ of an operation is \ATTN{categorical}, so we follow best practices and map it to a vector of \ATTN{parameters} called an \emph{opcode embedding}.
Opcode embeddings are concatenated with node features before being passed, along with the adjacency matrix, to a GNN. Kernel features are duplicated and concatenated with node feature vectors (`option 1' in \cref{fig:overview}).

\paragraph{Node Embedding}\label{sec:node-embedding}
We use a \ac{GNN} to combine information from the node and its neighbors to generate a node representation.
We use a \ac{GNN} because (i) a tensor computation kernel is naturally represented as a graph, and (ii) learning node representations conditioned only on their own features and local neighborhoods has shown to improve generalization in other settings.
We believe that local neighborhoods capture information that is important for estimating runtime.
For a example, node features include an output tensor shape but not input tensors' shape because operations can have variable numbers of inputs. With a \ac{GNN}, the model can receive input shape information from node's neighbors.

Our model builds on the GraphSAGE architecture~\cite{graphsage}.
We selected GraphSAGE since it is one of the simpler \ac{GNN} formulations that has been used successfully in inductive tasks.
The GraphSAGE embedding of node $i$ considering $k$-hop neighbors can be computed as follows:
\begin{equation*}
    \varepsilon^k_i = l_2\bigg(f^k_3\bigg(concat\Big(\varepsilon^{k-1}_i, \sum_{j \in \mathit{neighbors(i)}} f^k_2(\varepsilon^{k-1}_j \Big)\bigg)\bigg)
\end{equation*}
for all $k > 0$, and $\varepsilon^0_i = f_1(\OpsEmbed_i)$ otherwise.
Here: $f^k_{2 \dots 3}$ denote feedforward layers specific to depth $k$. $l_2$ denotes L2 normalization. $neighbors(i)$ is the set of immediate neighbors of node $i$. $\sum$ is a reduction chosen during hyperparameter search. %

\paragraph{Kernel Embedding \& Prediction}

We combine the node embeddings $\varepsilon^k$ to create the embedding $\kappa$ of the kernel.
We treat the exact method of calculating $\kappa$ as a hyperparameter, choosing from the following methods, including:
\setlist{nolistsep}
\begin{enumerate}
  \item a fully deterministic concatenation of one or more of column-wise maximum, mean, and/or sum reduction over $\varepsilon^k$ (\emph{column-wise} option),
  \item the final state of an LSTM~\cite{LSTM} on topologically sorted node embeddings, and
  \item the application of a Transformer encoder~\cite{Transformer} to the node embeddings.
\end{enumerate}
In each of these cases, the resulting kernel embedding will be linearly transformed into to scalar output by a feedforward layer without activation.
Additionally, we evaluate per-node predictions, which are the scalar outputs of a feedforward layer applied to each node embedding $\varepsilon_i^k$, and then we compute the sum of per-node predictions to get the kernel prediction (\emph{per-node} option).

The LSTM and Transformer reduction models are able to capture global and long-range dependency information, while the column-wise and per-node methods are not.

\subsection{Training Objectives}
\label{sec:task-adaptaion}

\paragraph{Tile-Size Selection Task}

In this task, we are interested in the relative speed between different tile sizes within each kernel.
Therefore, the performance model does not need to predict absolute runtime, but instead should be able to rank tile sizes by relative speed within each kernel. 
With this intuition, we train the model with a pairwise rank loss \cite{rank-loss-paper}: 
\begin{align}
    L &= \sum_{i = 1}^{n}\sum_{j = 1}^{n} \frac{\phi(y'_i - y'_j) \cdot pos(y_i - y_j)}{n \cdot (n-1) / 2}
    \label{eq:rank-loss}
\end{align}
where $n$ is the number of samples in each batch; $pos(z)$ is 1 if $z > 0$, or 0 otherwise; $\phi(z)$ is either the hinge function $(1 - z)_+$ or logistic function $log(1 + e^{-z})$, tuned via hyperparameter search.

\IGNORE{Alternatively, we can use the same MSE loss as in the fusion task, but weight a loss value of each sample appropriately so that the model is optimized for all kernels equally.}

\paragraph{Operator Fusion Task}

In this task, we would like the model to predict absolute kernel runtimes which can be used to compute total program runtime. Thus we minimize the model's squared error loss $(y'_i - y_i)^2$ with log-transformed targets. We apply log transformation because \ATTN{targets are right-skewed} and range from a nanosecond to a second.

\newcommand{\AppCount}{104}
\newcommand{\fusionTotalOps}{8580970471}
\newcommand{\fusionTotalKernels}{207361662}
\newcommand{\fusionTotalKernelsMils}{207}
\newcommand{\windowTotalOps}{1070224412}
\newcommand{\windowTotalKernels}{24376686}
\newcommand{\windowTotalKernelsMils}{24}

\section{Data}
\label{sec:dataset}

\begin{table*}[t]
    \centering
    \footnotesize
    \vspace{-.8em}
    \caption{The number of unique programs and kernels in the fusion and tile-size datasets. M = million.}
    \label{tab:dataset}

\begin{tabular}{lrrrrrrrr}
\toprule
{} & \multicolumn{4}{l}{\bf Random Split Method}
   & \multicolumn{4}{l}{\bf Manual Split Method} \\
\cmidrule(r){2-5}
\cmidrule(l){6-9}
{} & \multicolumn{2}{l}{\bf Programs}
   & \multicolumn{2}{l}{\bf Kernels}
   & \multicolumn{2}{l}{\bf Programs}
   & \multicolumn{2}{l}{\bf Kernels} \\
\cmidrule(r){2-3}
\cmidrule(lr){4-5}
\cmidrule(lr){6-7}
\cmidrule(l){8-9}
\textbf{Set        } &   Tile-Size & Fusion     &  Tile-Size &  Fusion
                    &   Tile-Size & Fusion     &  Tile-Size &  Fusion \\
\midrule
\textbf{Train}      &   93 &   78 &         21.8M  &       157.5M  &  93  & 78  & 22.9M &  190.2M \\
\textbf{Validation} &    8 &    8 &          1.6M  &        30.1M  &   8  &  8  &  1.4M &   11.2M \\
\textbf{Test}       &    8 &    8 &          1.4M  &        20.3M  &   6  &  6  &  0.5M &    6.6M \\
\bottomrule
\end{tabular}

    \vskip -0.1in
\end{table*}

Our dataset consists of \AppCount{} XLA programs used in production or commonly in research.
In order to test the ability of our approach to generalize to unseen programs, the programs were split into training, validation, and test sets in two ways:
(a) using the \emph{random} split method, 
in which programs were partitioned randomly into sets, and
(b) using the \emph{manual} split method, in which the test set was chosen by hand to minimize the subjective similarity of programs between the training and other two sets.
For each of the train, validation, and test sets, programs were expanded into individual kernels. 
\ATTN{\cref{tab:dataset} shows the number of programs and kernels in the training, validation, and test sets using both splitting methods.
The number of nodes per kernel is 41 on average across all programs, and ranges from 1 to 1,000.}
\ATTN{We measured the kernel runtimes on both TPUs v2 and v3. Our experiments use TPU v2 measurements unless mentioned otherwise.}

\paragraph{Tile-Size Dataset}
For the tile size dataset, we compiled each XLA program using the compiler's default fusion heuristics, obtaining an optimized computation graph that we decompose into kernels.
For each kernel, we queried the compiler for a list of valid tile sizes.
The runtime target for each sample is the minimum runtime from three runs.
A kernel may have as many as 500,000 valid tile sizes, so we measured runtimes for as many as possible for each kernel within 30 minutes across 50 hosts, each with an accelerator. \ATTN{This process generated the total of 25 million samples.}

\paragraph{Fusion Dataset}
For the fusion dataset, we ran the fusion autotuner with a random search strategy to generate, for each computation graph, 50,000 fusion configurations or until timeout (4 hours using 50 machines).
Graphs were then decomposed into kernels, %
\ATTN{yielding 208 million samples after duplicate elimination.}
Approximately half of the resulting kernels have runtimes below 5$\mu$s.
These contribute negligibly to total program runtimes, so we emphasize larger kernels in our analysis.

\paragraph{Imbalances}
Our data are imbalanced in two ways.
First, programs are not wholly independent. For example, there are many variations of ResNet models, but just one AlexNet model and one DLRM (recommendation) model.
Second, the number of kernels and tile sizes vary widely across different models. In the fusion dataset, ResNet variant models have 300x more samples than AlexNet variants, and in the tile-size dataset, models using Inception have 400x more kernels than auto-completion models.
To account for these imbalances, we draw examples evenly from each model type during training.

\section{Model Accuracy Evaluation}
\label{sec:accuracy}

We trained our models on a single NVidia V100 instance with 96GB of RAM and 10 CPU cores \CUT{for data processing}.
For all the learned models, we did a hyperparameter search (presented in \cref{sec:hp}) and selected the best-performing model for each task on the validation set.

\begin{table*}[t]
    \centering
    \footnotesize
    \vspace{-.5em}
    \caption{
        The main evaluation metrics for both tasks on the randomly split test set, grouped by test application, comparing our best learned performance models against the analytical baseline.
        Geometric mean and median statistics are over application-level metrics.
        Fusion experiment statistics are evaluated over kernels with $\geq$5$\mu$s true runtimes, which account for the majority of total runtime in our programs. 
    }
    \label{tab:fusion-and-window-per-app}
    
\begin{tabular}{@{}lrrrrrrrr@{}}
\toprule
{} & \multicolumn{4}{c}{\textbf{Tile-Size}}
   & \multicolumn{4}{c}{\textbf{Fusion}} \\
\cmidrule(r){2-5}
\cmidrule(l){6-9}
{} & \multicolumn{2}{l}{\textbf{Tile-Size APE}}
   & \multicolumn{2}{l}{\textbf{Kendall's $\tau$}}
   & \multicolumn{2}{l}{\textbf{MAPE}}
   & \multicolumn{2}{l}{\textbf{Kendall's $\tau$}} \\
\cmidrule(r){2-3}
\cmidrule(lr){4-5}
\cmidrule(lr){6-7}
\cmidrule(l){8-9}
{} & \textbf{Learned}
   & \textbf{Analytical}
   & \textbf{Learned}
   & \textbf{Analytical}
   & \textbf{Learned}
   & \textbf{Analytical}
   & \textbf{Learned}
   & \textbf{Analytical} \\
\midrule
\textbf{ConvDRAW}   &   9.7 &   3.9 &  0.75 &   0.79   &   17.5 &    21.6 &   0.80 &   0.77 \\
\textbf{WaveRNN}    &   1.5 &   2.8 &  0.75 &   0.65   &    2.9 &   322.9 &   0.97 &   0.70 \\
\textbf{NMT Model}  &   3.1 &  13.1 &  0.86 &   0.81   &    9.8 &    26.3 &   0.94 &   0.91 \\
\textbf{SSD}        &   3.9 &   7.3 &  0.82 &   0.77   &   11.4 &    55.9 &   0.88 &   0.76 \\
\textbf{RNN}        &   8.0 &  10.2 &  0.64 &   0.55   &    1.9 &    20.5 &   0.97 &   0.86 \\
\textbf{ResNet v1}  &   2.8 &   4.6 &  0.85 &   0.73   &    3.1 &    11.5 &   0.95 &   0.88 \\
\textbf{ResNet v2}  &   2.7 &   5.4 &  0.87 &   0.73   &    2.4 &    13.3 &   0.96 &   0.86 \\
\textbf{Translate}  &   3.4 &   7.1 &  0.93 &   0.92   &    2.1 &    27.2 &   0.92 &   0.74 \\
\midrule

\textbf{Median}     &   3.3 &   6.2 &  0.84 &   0.75 &    3.0 &    24.0 &     0.95 &   0.82   \\
\textbf{Mean}       &   3.7 &   6.1 &  0.80 &   0.74 &    4.5 &    31.1 &     0.92 &   0.80   \\
\bottomrule
\end{tabular}

    \vskip -0.1in
\end{table*}

\subsection{Tile-Size Task}
\label{sec:eval-tile-size}

\paragraph{Metrics}

For this task, we are interested in relative runtimes between different tile sizes within each kernel.
Thus, for each kernel, we find the tile size with the best predicted runtime and the one with the best true runtime, and find the difference between their \emph{true} runtimes.
This is distinct from measuring differences between predicted runtimes and true runtimes.
The `Tile-Size APE' (listed in \cref{tab:fusion-and-window-per-app}) is computed by summing the differences across all program kernels and dividing the sum by the runtime of the program as if it had chosen the best tile size for every kernel.
More precisely, the Tile-Size APE of a program with kernels $k \in K$ and set of tile size configurations $C_k$ is:
\begin{equation}%
    100 \times \frac{
        \sum_{k \in K} |t_{c'_k}^k - \min_{c \in C_k} t_c^k|
    }{
        \sum_{k \in K} \min_{c \in C_k} t_c^k
    }
\end{equation}%
where $t_c^k$ is the true runtime of tile size configuration $c$ for kernel $k$, and $c'_k$ is the predicted-best configuration.
This is a good measure of efficacy for the setting, in which we use the performance model to select the top candidates and verify their actual runtimes using real hardware. Tile-Size APE shows how far we are from the fastest program.
We also measure the Kendall rank correlation between targets and predictions of tile-size runtimes within each kernel, and compute the average over all kernels in each program.

\paragraph{Results}
\Cref{tab:fusion-and-window-per-app} shows results for the randomly split dataset. The baseline is XLA's mature analytical performance model designed for this task, 
as described in \cref{sec:manual-cost}. Our learned performance model (3.7\% mean error and 0.8 mean correlation) performs slightly better than the analytical model (6.1\% mean error and 0.74 mean correlation). Our learned model is consistently better than the analytical model on all benchmarks except ConvDraw, which differs more (subjectively) from the programs in our training set than any other test program. \ATTN{TPU v3 results are similar; the learned performance model has 3.8\% mean error with a slightly lower mean correlation of 0.65.}
\IGNORE{This is not surprising due to the fact that the training set contains samples from only one ConvDraw program (with a different batch size), whereas the training set contains more diverse samples similar to the other test benchmarks.}

On the manually split dataset, the learned model (6.3\% mean error) performs slightly worse than the analytical model (2.3\% mean error).
It is expected that the test error of the learned model on this test set will be higher than that of the randomly split test set, as these test programs were chosen for their dissimilarity to the training set.
\ATTN{See \cref{tab:fusion-and-window-per-app-manual-split} in the appendix for more detail.}

\subsection{Fusion Task}
\label{sec:eval-fusion}

\paragraph{Metric}
In this task, we use \ac{MAPE} as we wish to estimate the absolute runtime of the kernels in order to predict the total program runtime.

\paragraph{Baseline}
The existing analytical performance model in XLA is built for selecting the fastest tile size given a kernel, so performance estimates for different kernel types (e.g., fused kernels with and without convolutions) are in different scales. 
Hence, we scale the analytical model's output with a coefficient associated with the kernel's type to get an estimated absolute runtime.
Coefficients are determined by executing each program in the test set with a default fusion configuration, and dividing the actual total runtime for all kernels of each type by the estimate in its original scale. The analytical model does not support kernels without tile-size options, which account for 1\% of kernels in the dataset. We ignore these kernels in our comparisons in this section.

\paragraph{Results}
\Cref{tab:fusion-and-window-per-app} reports MAPEs of kernels with runtimes $\geq$5$\mu$s.
Our best learned model (4.5 \ac{MAPE} and 0.92 mean correlation) substantially outperforms the analytical model (31.1 MAPE and 0.8 mean correlation).
Similar to the tile-size dataset, our model consistently performs better than the analytical model on all benchmarks.
On kernels with $<$5$\mu$s runtimes, results follow the same trend; our model and the analytical model have \ac{MAPE}s of 5.0 and 22.7; and mean Kendall's $\tau$ coefficients of .89 and .7 respectively.
\ATTN{For the kernels with runtimes $\geq$5$\mu$s on TPU v3, the learned performance model has 4.9 \ac{MAPE} and 0.92 mean correlation.}

On the harder manual split, the learned model still outperforms the analytical model significantly \ATTN{(see \cref{tab:fusion-and-window-per-app-manual-split} in the appendix)}.
On kernels with runtimes $\geq$5$\mu$s, our model and the analytical model have \ac{MAPE}s of 6.2 and 18.1 respectively.

\section{Model Ablation Studies}
\label{sec:ablation}

We ran a comprehensive set of ablation experiments to study the effects of design decisions underlying the best-performing model presented in \cref{sec:accuracy}, including the objectives used, the presence of static performance features, and the model architecture.
Experiments in this section use the randomly split datasets and the same evaluation metrics as in the previous section: Tile-Size APE for the tile-size task and \ac{MAPE} for the fusion task.
Each ablation (row in \cref{tab:ablation-features}) is a single change to the `vanilla' configuration.

\subsection{Graph Features and Loss Function}
\label{sec:features-and-loss}

To determine what input features are important and the suitable training loss function to use, we used the same neural network model across all the experiments in \cref{sec:features-and-loss}.
In particular, we used GraphSAGE with the simple per-node reduction, which is quick to train, and one of our best-performing hyperparameter configurations.
Each model configuration was trained for 3 million steps.

\paragraph{Edge Direction}
First, we considered a model variant that, unlike the `vanilla' model,
\ATTN{applied the same feedforward network to node representations from incoming and outgoing edges (see `Undirected' in \cref{tab:ablation-features}).}
The results suggest that edge direction is important for the fusion task---reducing the mean error by 3.8\%---but irrelevant to the tile-size task. 

\paragraph{Static Performance Features}
The `With static perf. (as node features)' row of \cref{tab:ablation-features} shows the result when we add four static performance features---as explained in \cref{sec:model-inputs}---to the `vanilla' model, which uses only features that are extracted directly from the XLA program representation. Similar to edge directions, these features significantly improve model accuracy for the fusion task---reducing the mean error by 5\%---but less so for the tile-size task. 

The finding that edge direction and static performance information help only the fusion task is somewhat unexpected but not entirely surprising. In the tile size selection task, we predict the relative runtimes of different tile sizes of the same kernel, but never compare runtimes of different kernels. Thus, the static performance features and the kernel graph are constant across different tile sizes, and the only changing input features are the tile size features. However, these constant features may still help determine the relative runtimes of different tile sizes more accurately, as we can see that the static performance features slightly improve the tile size runtime prediction accuracy. Hence, adding more input features may not help significantly if they are constant across kernels that will be compared against each other.

\begin{table}[t]
    \centering
    \footnotesize
    \vspace{-.8em}
    \caption{The table reports Tile-Size APE for the tile-size dataset, and MAPE for the fusion dataset, on test programs for a variety of model variants. All models are trained for 3 million steps.
    \textdagger~The model configuration used in \cref{sec:accuracy}.}
    
\begin{tabular}{lrrrr}
\toprule
{} & \multicolumn{2}{l}{\textbf{Tile-Size}} & \multicolumn{2}{l}{\textbf{Fusion}} \\
\cmidrule(r){2-3}
\cmidrule(l){4-5}
{} &     Median & Mean &                  Median & Mean  \\
\midrule
\textbf{Vanilla}                &   6.2 &  6.8  &  9.5 &  10.2  \\
\textbf{Undirected}       &   7.2 &  6.8  & 11.0 &  14.0  \\
\textbf{With static perf. \textdagger}   &   6.5 &  6.3  &  4.0 &   5.2  \\
(as node features)     &       &       &      &        \\
\textbf{With static perf.}   &   6.1 &  5.9  &  5.7 &   6.0  \\
(in kernel embedding)   &       &       &      &        \\
\textbf{Move tile-size} (node    &  10.2 &  9.4  &  N/A &    N/A \\
 feats. to kernel emb.)   &       &       &      &        \\
\textbf{MSE loss} (not rank)         &  16.7 & 17.7  & N/A  &    N/A \\
\bottomrule
\end{tabular}
    \vspace{-1em}
    \label{tab:ablation-features}
\end{table}

\begin{table*}[t]
    \centering
    \footnotesize
    \vspace{-.8em}
    \caption{Model ablation study results. The table reports Tile-Size APE for the tile-size dataset, and MAPE for the fusion dataset, on test programs. Standard deviations of the errors across test applications are in parentheses.
    Reductions (rows) are defined in \cref{sec:model-design}.
    \textbf{Bold} indicates the selected models used in \cref{sec:accuracy}.
    All models are trained until 5 million steps. }

\begin{tabular}{lrrrrrr}
\toprule
{} & \multicolumn{3}{l}{\textbf{Tile-Size}}
   & \multicolumn{3}{l}{\textbf{Fusion}} \\
\cmidrule(lr){2-4}
\cmidrule(l){5-7}
{\bf Reduction $\setminus$ Graph}
  & No GNN
  & GraphSAGE
  & GAT
  & No GNN
  & GraphSAGE
  & GAT  \\
\midrule
per-node     &    10.7 (5.3) & 6.0 (3.8)          & 9.2 (6.4)  & 16.6 (132.7)   &  7.3 (34.6)      & 15.1 (4.0) \\
column-wise  &    9.3 (3.3)  & 6.9 (3.0)          & 8.4 (4.2)  &  6.6 (9.1)     &  5.1 (3.6)       & 8.5 (3.8) \\
LSTM         &    7.1 (3.7)  & \textbf{3.7 (2.8)} & 7.7 (4.2)  &  3.9 (7.5)   &  5.0 (4.3) & 7.4 (4.5) \\
Transformer  &    10.8 (7.4) & 4.6 (2.6)          & 8.2 (3.8)  &  7.3 (10.1)   &  \textbf{4.5 (5.8)}        & 14.6 (11.3)\\
\bottomrule
\end{tabular}
    \vspace{-1em}
    \label{tab:ablation-models}
\end{table*}

\paragraph{Kernel Features Encoding}
Two ways to encode kernel information are shown in \cref{fig:overview}, labeled `kernel features (option 1)' and `(option 2)'.
In \cref{tab:ablation-features}, the `vanilla' model uses option 1, whereas the `Move tile-size (node feats. to kernel emb.)' uses option 2. Encoding tile-size with node features outperforms encoding it with the kernel embedding (2.6\% lower mean error). We believe this is because tile size can be important for estimating runtime for an individual operation before aggregation. \ATTN{When the tile-size information is available at the node level, the model still has all the information about the node, such as its input and output shapes.}
On the other hand, encoding static performance information as node features or kernel features makes little difference because these features are not very important for estimating the runtime for an individual node.

\paragraph{MSE vs. Rank Loss}

For the tile-size dataset, we compare using MSE and rank loss as the training objective.
The `vanilla' model is trained using rank loss, while the `MSE loss (not rank)' in \cref{tab:ablation-features} uses MSE loss. The effect is significant, using rank loss is 10.9\% more accurate.
This result confirms our intuition that training a model to predict relative speeds is easier than absolute runtimes.

\subsection{Neural Network Model}

Once we determined the best set of graph features to include and the suitable loss function for each task from the experiment in \cref{sec:features-and-loss}, we performed a comprehensive comparison between different modeling choices \ATTN{to answer:}
\begin{enumerate}
\itemsep0em 
    \item Does a GNN outperform models that treat programs as sequences?
    \item Do we need an additional model to capture long-range dependencies in a kernel beyond a GNN?
    \item How does GraphSAGE compare to a more sophisticated GNN: \ac{GAT}?
\end{enumerate}

To answer these questions, we explore different combinations of modeling choices for a GNN (GraphSAGE, GAT, and no GNN) and node reduction methods (per-node, column-wise, LSTM, and Transformer).

For all the models in this experiment, we train to five million steps and use the best settings from \cref{sec:features-and-loss}: distinguishing edge direction, and including static performance information (and tile-size) as node features. 

\paragraph{Q1: Graphs vs. Sequences}
Prior work proposes an LSTM-based performance model for x86 basic blocks~\cite{ithemal}.
To understand the effect of representing program examples as graphs rather than sequences, we compare GraphSAGE with the simple column-wise reduction to LSTM and Transformer (with no GNN). LSTM and Transformer models are trained over topologically sorted sequences of nodes, whose embeddings are the same per-node representations fed into GNNs.

According to \cref{tab:ablation-models}, GraphSAGE with the column-wise reduction is more accurate than using LSTM or Transformer without a GNN on the tile-size dataset. On the fusion dataset, LSTM is slightly better than GraphSAGE, but LSTM has a higher error variance across test applications. Since we want the performance model to be consistently accurate across all programs, we conclude that GraphSAGE is a crucial component to achieve that.
Another interesting finding is that Transformer alone is worse than the simple column-wise reduction even without GraphSAGE.

\paragraph{Q2: Most Effective Global Reduction}
GNNs capture local graph structural information but not the global structure. To consider some global information and long-range dependencies in a kernel, we apply a sequence model (LSTM) and a global attention model (Transformer) to generate kernel embeddings from node embeddings produced by a GNN.

As seen in \cref{tab:ablation-models}, applying either LSTM or Transformer on top of GraphSAGE improves the model accuracy over GraphSAGE with a non-model-based reduction (per-node or column-wise). This result suggests that in order to achieve the best accuracy, the model indeed needs to capture more than local dependencies. GraphSAGE-LSTM and GraphSAGE-Transformer perform equally well on both tile-size and fusion datasets. However, GraphSAGE-Transformer is much faster to train.

Nevertheless, GraphSAGE with the simple column-wise reduction already works reasonably well. If one prefers a model with fast inference time, we recommend such a combination. While GraphSAGE with the per-node reduction is slightly better than GraphSAGE with the column-wise reduction on the tile-size task, it is significantly worse on the fusion task with a high variance across applications.

\paragraph{Q3: Most Effective GNN}
We compared our choice of GraphSAGE to \ac{GAT}, which found state-of-the-art performance on a number of benchmarks~\cite{gtn}.
We use \ac{GAT} with multiple attention heads per each layer.
According to Table~\ref{tab:ablation-models}, GraphSAGE consistently exhibits better test accuracy compared to GAT.
We noticed that training GATs was especially sensitive to hyperparameter choices.
For example, GraphSAGE was less sensitive to learning rate changes than GATs. Therefore, we conclude that GraphSAGE with roughly the same number of learnable parameters compared to GAT generalizes better for our cost prediction regression task. Additionally, we observed that GAT with LSTM/Transformer is worse than LSTM/Transformer alone. We hypothesize that training a compounded complex model further increases the training difficulty.

\section{XLA Toolchain Integration}
\label{sec:eval-autotune}

In this section, we integrated the learned model into the XLA compiler and autotuner.

\subsection{Tile-Size Compiler Integration}

We integrated the model directly into the XLA compiler, replacing the analytical model. \cref{fig:autotune-window}'s `Learned model 1' shows the benchmark speedup over the default tile-size configurations (the best tile-sizes according to the analytical cost model). 
The first eight benchmarks are from the test set, and the remaining four are benchmarks that gain most speedup from exhaustive search. 

On the test set benchmarks, the learned model is comparable to the analytical model, except for ConvDraw. We observe a few percent slowdown on NMT, SSD, and Translate even though our model shows better accuracy on these benchmarks in \cref{tab:fusion-and-window-per-app}. This is likely because the dataset does not contain all possible tile sizes for a kernel if the time limit is reached during data generation.

On 3 out of 4 additional benchmarks, the learned cost model is better than the analytical model. On Translate (3), replacing the compiler's analytical model with the learned model would yield a 20\% speedup.
This demonstrates another advantage of a learned performance model over a manually-written model: it can be easily improved with more data. If the learned model does not perform well on some benchmarks, we can re-train or fine-tune the model on similar benchmarks. In contrast, to fix the analytical model, engineers must identify the problem and fix it in a way that does not hurt other benchmarks, which is challenging in practice.

\subsection{Tile-Size Autotuner Integration}
\label{sec:eval-autotune-window}

\begin{figure}[t]%
    \includegraphics[width=\linewidth]{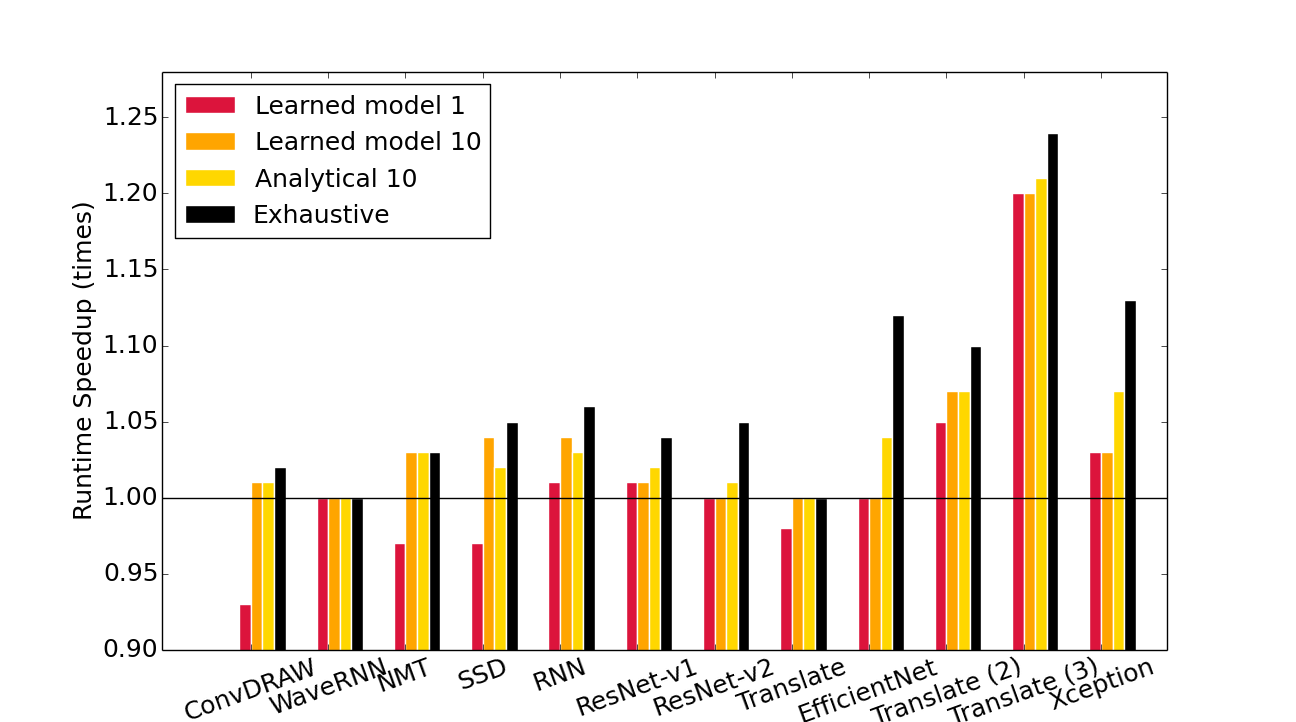}
    \vspace{-2em}
    \caption{Runtime speedup, compared to the default tile size configurations, found by the tile-size autotuner using the learned model (`Learned model 10') and analytical model (`Analytical 10') to select the top 10 candidates per kernel to run on real hardware. \ATTN{A speedup of 1 means performance matches the standard analytical model.} `Exhaustive' reports the speedup found by an exhaustive search. `Learned model 1' shows speedup when replacing the analytical model with the learned model directly in the compiler.}
    \vspace{-1em}
    \label{fig:autotune-window}
\end{figure}

Instead of using the learned model inside the compiler directly,
we can use it in the tile-size autotuner. 
\cref{fig:autotune-window} reports the end-to-end benchmark speedup found by the autotuner.
By default the autotuner enumerates all tile-sizes for each kernel, and evaluates them on hardware (labeled `Exhaustive'). Instead, we use the learned performance model (labeled `Learned model 10') and the analytical model (labeled `Analytical 10') to select the top 10 tile-sizes to be evaluated on hardware as a way to reduce the search time. The figure shows that the learned model is on par with the analytical model across all benchmarks (within 1-3\% of each other).

\subsection{Fusion Autotuner Integration}

We also integrate the best learned performance model from \cref{sec:eval-fusion} in the XLA fusion autotuner.
The analytical model is not used in this experiment as it cannot estimate runtimes for kernels that lack tile-size options; kernels that are not fusion, convolution, or data formatting operations.

\paragraph{Experimental Setup}
TPUs are in high demand, so we wish to minimize their use during autotuning.
CPUs are more abundant and better support time-sharing, and, with a performance model, can be used to more cheaply run the autotuner.
We compare the baseline autotuner (which uses TPUs) with the learned autotuner (which uses both the learned performance model and a \ac{TPU}).
In this experiment, the autotuner searches via simulated annealing.
The baseline autotuner evaluates fusion configurations on real hardware for 10 minutes. The learned autotuner first evaluates fusion configurations on a CPU for an hour, then runs promising fusion configurations on the real hardware for up to either 1 or 10 minutes, in the order ranked by the predicted costs.

In this experiment, we compare the fusion autotuner on a set of programs that gain significant speedup from autotuning. \ATTN{The autotuner starts the search from a default configuration, which is generated by the compiler's fusion heuristic given a specific program.}
Although some test programs (Transformer, Char2Feats, and ResNet-parallel) are in our training set,
most kernels seen during the evaluation are unlikely included in the training set.
This is because kernels in the training set are generated using a random search as opposed to the simulated annealing used during this evaluation; as a result, different kernels are produced even for the same program.

\paragraph{Results}
We run the autotuner on each program 10 times and report the best speedup found over the default configuration in
\cref{fig:autotune-fusion-default}.
Using the learned performance model together with the hardware let us discover fusion configurations that are on average 1.5\% faster than using the hardware alone. Additionally, they are on average only 1.5\% slower than the best known configurations found when running the autotuner on hardware for 4 hours.
When running simulated annealing starting from a random configuration, the benefit from the performance model is even more pronounced. On average, using the performance model led to discovering 10\% faster configurations \ATTN{compared to not using the performance model}. %

Furthermore, the learned performance model reduces time spent using real target hardware for evaluation from 10 minutes to 1 minute without degrading performance. \ATTN{This demonstrates that when access to a target hardware is limited, the autotuner can utilize the learned performance model to discover faster code.}
This experiment shows that our approach can indeed be used to build a practical, accurate performance model to guide a compiler optimization task.

\IGNORE{
\begin{figure}[t]%
    \begin{subfigure}[t]{1.\linewidth}
      \includegraphics[width=\linewidth]{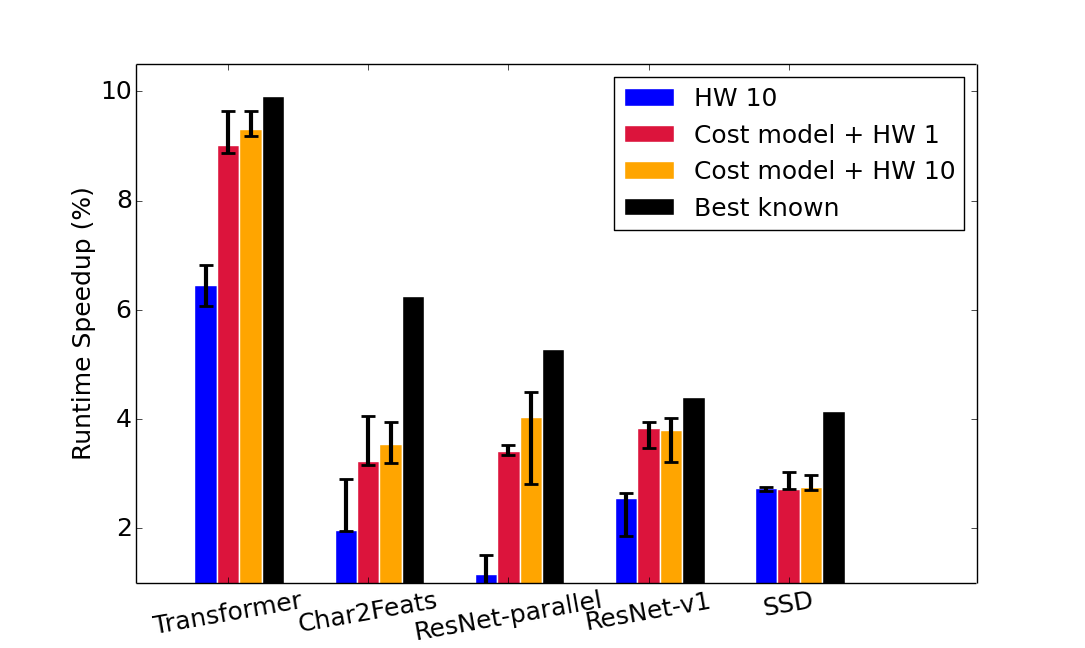}
      \caption{Autotuning from default configuration}
    \label{fig:autotune-fusion-default}
    \end{subfigure}
    \hfill
    \begin{subfigure}[t]{1.\linewidth}
      \includegraphics[width=\linewidth]{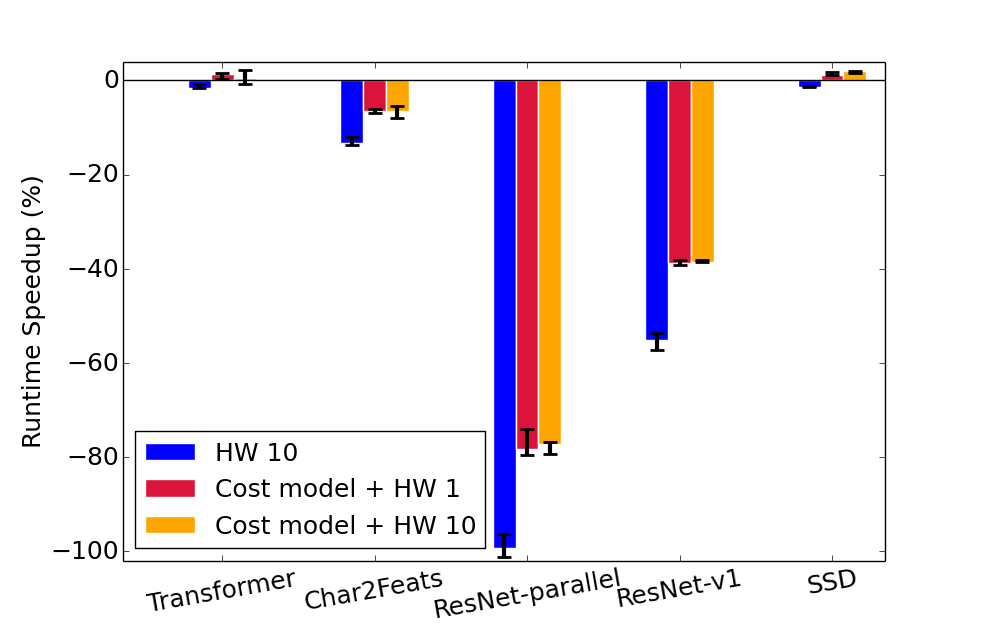}
      \caption{Autotuning from random configuration}
    \label{fig:autotune-fusion-random}
    \end{subfigure}
    
    \caption{Runtime speedup obtained by the fusion autotuner using the hardware alone (HW) or using the learned performance model with hardware (Cost model + HW). Speedup is computed over the default compiler-chosen configuration. The $m$ in the label `HW $m$' indicates the time duration of using the hardware in minutes. We ran the experiment three times. The solid bars show the median best speedup with the error bars indicating min and max.}
    \vspace{-1em}
    \label{fig:autotune-fusion}
\end{figure}
}

\begin{figure}[t]
    \includegraphics[width=\linewidth]{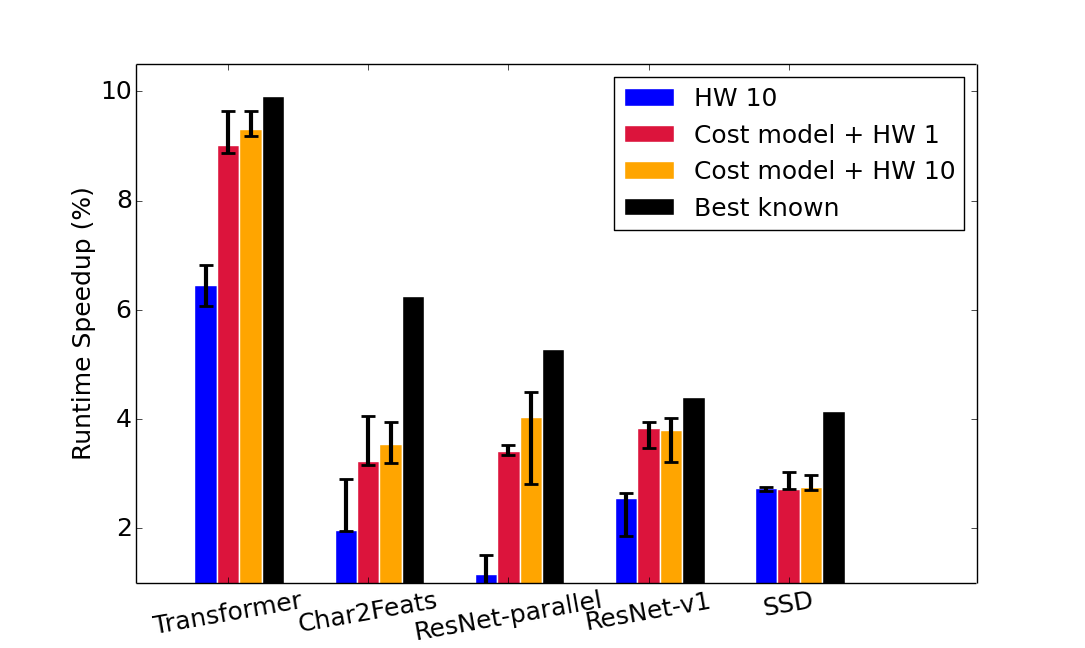}
    \vspace{-2em}
    \caption{Runtime speedup obtained by the fusion autotuner using hardware alone (HW) or using the learned performance model with hardware (Cost model + HW). Speedup is computed over the default compiler-chosen configuration. The $m$ in the label `HW $m$' indicates minutes using the hardware. We ran the experiment 3 times. Solid bars show the median best speedup; error bars range from min to max.}
    \vspace{-1em}
    \label{fig:autotune-fusion-default}
\end{figure}

\section{Related Work}
\IGNORE{
Many researchers have applied machine learning to code optimization. Here, we consider efforts that use machine learning to build performance models to predict the program's execution time.}

Ithemal uses a hierarchical recurrent neural network to estimate throughputs of x86-64 basic blocks~\cite{ithemal}.
Basic blocks are short,
loop-free sequences of instructions (6.06 instructions on average). In contrast, our work addresses larger kernels with implicit nested loops 
containing up to a thousand operators.
Ithemal was evaluated on its ability to generalize to held-out basic blocks. However, our method is tested for its ability to generalize to novel tensor programs and targets a very different processor.

The code feature-based performance model~\cite{code-features-based}, Halide's performance model~\cite{autohalide}, and work by Justus et al.~\cite{justus-et-al-model} use simple neural networks to predict runtime from manually-engineered features produced by a static analyzer that examines
an optimized program.
Since extracting these features from an XLA graph is non-trivial, we train a more complex neural net---using features that can be extracted directly from the XLA graph and very minimal features produced by an already existing static analyzer---with sufficient capacity to recover similarly powerful representations.

AutoTVM also uses a learned performance model to optimize tensor kernels, by ranking candidates~\cite{AutoTVM}.
However, AutoTVM's model shows limited ability to generalize between even very similar individual kernels (e.g., different kinds of convolution).
In contrast, we train a performance model over entire tensor programs with many kernels, and can generalize to novel tensor programs containing many kernels dissimilar to those seen during training.

Additionally, \ac{NAS} often employs a related idea: learning models to predict the error of an deep learning model architecture (e.g, \cite{peephole, istrate2018tapas, wen2019neural}).
Others, such as ReNAS~\cite{xu2019renasrelativistic}, learn to rank candidate neural architectures rather than predict runtimes in isolation.

Deep learning-based techniques have been proposed to find better compiler optimizations~\cite{end-to-end-deeplearning, Chameleon}. More specifically, GNNs have been used in the context of various compiler optimizations. ProGraML~\cite{programl} uses GNNs to perform compiler analyses. Vemal~\cite{vemal} proposes imitation learning-based auto-vectorization based on gated GNNs. Reinforcement learning- and evolutionary search-based techniques using GNN-based policies have been proposed for the device placement task~\cite{regal,placeto, zhou2019gdp}. %

\section{Conclusion}

We have presented an approach for automatically learning a performance model for tensor programs.
\ATTN{We have found that the learned model %
can generalize well to programs with some similarity to our training set, usually matching or improving upon the performance of the best known analytical baseline for our target hardware.} %
\ATTN{We also} demonstrated that the learned performance model can be employed by autotuners to discover faster tensor programs than using hardware targets alone when hardware access is limited.
\ATTN{In addition,} we showed several advantages of the learned approach over the manual one, beyond accuracy. First, we have created, without manual feature engineering, a new performance model for the XLA fusion task where none existed before. Second, we can improve the learned model by re-training or fine-tuning with more data. 
\IGNORE{However, the learned approach creates a neural network black box that is incomprehensible to humans. Machine learning explainability is an active area of research that we think can be applied to learned performance models. Another direction for future work is to use the combination of neural network and program synthesis to generate performance models in the form of programs. }

\section*{Acknowledgements}

We would like to thank the XLA team, especially Bjarke Roune, for feedback and help during development, Hyeontaek Lim for code review and writing feedback, and Rishabh Singh for guidance.

\bibliography{refs}
\bibliographystyle{mlsys2021}

\clearpage
\appendix

\section{Tile-Size Analytical Model}
\label{appendix:manual-model}

A key to achieving high performance is to use the fast scratchpad memory effectively.
Choosing an appropriate tile size is essential to achieving this goal for a number of reasons:
\begin{enumerate}
\itemsep0em 
  \item Tile selection determines the number of times data has to be copied between the HBM and scratchpad memory. A bad tile choice may result in a larger data movement.
  \item Tile selection determines the quantity of data that gets copied in a given iteration, and the amount of compute performed in that iteration. A good balance between the two is essential for achieving high performance through the overlap of compute and data transfers.
  \item Because tile size determines the data size copied, it also controls the achieved bandwidth for data transfers. Larger transfers are more efficient.
\end{enumerate}

The analytical model estimates the kernel's data transfer time and computation time, taking the maximum of the two. The compiler pipelines the code overlapping computation of a given tile with the data copy-in (HBM to scratchpad) of the next tile, and data copy-out (scratchpad to HBM) of the previous tile. The performance model takes into account the memory required by all the operations it contains. The compiler's code generation scheme distributes operations among the functional units while respecting data dependencies. To estimate the computation cost, the model must estimate the instruction schedule for each operation to determine the critical path. 
Since different tiles may demand and execute different amounts of data transfer and computation, the total cost is determined on a per-iteration basis.

Since the tile-size selection happens prior to the code generation, it has to rely on several heuristics due to: (i) inability to accurately estimate bi-directional data transfers, (ii) limitations in modeling instruction scheduling, (iii) inability to model the effect of register usage, and (iv) limitations in capturing dynamic execution properties, such as issue stalls. The heuristics are chosen by tuning the performance model on a set of benchmark programs. 

\section{Hyperparameters}
\label{sec:hp}

\Cref{tab:fixed-hp} shows the fixed hyperparameters we used in all experiments. These hyperparameters were tuned in prior preliminary experiments.
\Cref{tab:hp-tile-size} and \cref{tab:hp-fusion} reports the hyperparameters of the best performing models in \cref{tab:ablation-models} for the tile-size and fusion datasets, respectively.

The model hyperparameters used to produce \cref{tab:ablation-features} are the same as `GraphSAGE + per-node' in \cref{tab:hp-tile-size} and \cref{tab:hp-fusion}. The training hyperparameters are slightly different but in the same range as we always tuned these parameters in every experiment.

\begin{table*}[t]
    \centering
    \footnotesize
\begin{tabular}{lll}
\toprule
Hyperparameter & Applicable to & Fixed value  \\
\midrule
Opcode embedding size     & All    & 256 \\
Node neighbor size $^a$   & GNN    & 20  \\
GNN layers           & GNN    & 3   \\
GraphSAGE aggregator & GNN    & mean   \\
Node final layers $^b$    & All    & 3   \\
Column-wise reduction type  & Column-wise reduction & mean \& max $^c$   \\
Transformer attn. heads    & Transformer    & 4   \\
Transformer reduction      & Transformer    & sum   \\
Include per-layer biases   & All            & no   \\
\bottomrule
\end{tabular}
    \caption{Fixed model's hyperparameters used in all experiments. \\
    $^a$ Node neighbor size is the maximum number of neighbors included per node. If a node has more neighbors, we truncate the neighbor list. We experiment with sampling instead of truncation, but there is no difference. \\
    $^b$ Node final layers is the number of feedforward layers applied to node embeddings before reduction. \\
    $^c$ Concatenation of column-wise mean and column-wise max. \\
    }
    \label{tab:fixed-hp}
\end{table*}

\begin{table*}[t]
    \centering
    \scriptsize
\begin{tabular}{p{2.8cm}|p{0.7cm}p{1cm}p{0.7cm}p{1cm}p{0.7cm}|p{1cm}p{1cm}p{0.7cm}p{0.7cm}p{0.7cm}}
\toprule
& \multicolumn{5}{c|}{\textbf{Model Hyperparameters}}& \multicolumn{5}{c}{\textbf{Training Hyperparameters}} \\
\textbf{Tile-size dataset} & Hidden dim. & Module L2 norm & Module layers & Transformer layers & GAT head & Learning rate  & Learning rate decay & Grad. clip & Dropout & Rank loss \\
\midrule
No GNN + per-node & 512 & False & 3 & N/A & N/A & 0.000802 & 1.0 & none & 0.1 & hinge \\
No GNN + column-wise & 1024 & False & 3 & N/A & N/A & 0.000642 & 1.0 & none & 0.1 & hinge \\
No GNN + LSTM & 512 & False & 0 & N/A & N/A & 0.000434 & 0.99 & norm & 0.1 & hinge \\
No GNN + Transformer & 1024 & False & 0 & 3 & N/A & 0.000424 & 0.99 & norm & 0.1 & hinge \\
\midrule
GraphSAGE + per-node & 512 & False & 0 & N/A & N/A & 0.001526 & 1.0 & none & 0.1 & hinge \\
GraphSAGE + column-wise & 1024 & False & 0 & N/A & N/A & 0.000642 & 1.0 & none & 0.1 & logistic \\
GraphSAGE + LSTM & 1024 & False & 0 & N/A & N/A & 0.000386 & 0.98 & norm & 0.1 & hinge \\
GraphSAGE + Transformer & 1024 & False & 0 & 1 & N/A & 0.000466 & 1.0 & norm & 0.1 & hinge \\
\midrule
GAT + per-node & 512 & False & 0 & N/A & 2 & 0.00001 & 1.00 & norm & 0.1 & hinge \\
GAT + column-wise & 512 & False & 0 & N/A & 4 & 0.00001 & 1.00 & norm & 0.1 & hinge\\ 
GAT + LSTM & 512 & False & 0 & N/A & 4 & 0.00001 & 0.99 & norm & 0.1 & hinge\\
GAT + Transformer & 512 & False & 0 & 2 & 4 & 0.00001 & 1.00 & norm & 0.1 & hinge\\
\bottomrule
\end{tabular}
    \caption{Tuned hyperparameters used in the model ablation experiment on the tile-size dataset.}
    \label{tab:hp-tile-size}
\end{table*}

\begin{table*}[t]
    \centering
    \scriptsize
\begin{tabular}{p{2.8cm}|p{0.7cm}p{1cm}p{0.7cm}p{1cm}p{0.7cm}|p{1cm}p{1cm}p{0.7cm}p{0.7cm}}
\toprule
& \multicolumn{5}{c|}{\textbf{Model Hyperparameters}}& \multicolumn{4}{c}{\textbf{Training Hyperparameters}} \\
\textbf{Fusion dataset} & Hidden dim. & Module L2 norm & Module layers & Transformer layers & GAT heads &  Learning rate & Learning rate decay & Grad. clip & Dropout \\
\midrule
No GNN + per-node & 512 & False & 3 & N/A & N/A & 0.000214 & 0.95 & none & 0.2 \\
No GNN + column-wise & 512 & False & 3 & N/A & N/A & 0.000102 & 1.0 & none & 0.25 \\
No GNN + LSTM & 512 & False & 0 & N/A & N/A & 0.000144 & 1.0 & none & 0.25 \\
No GNN + Transformer & 512 & True & 0 & 1 & N/A & 0.000862 & 1.0 & norm & 0.25 \\
\midrule
GraphSAGE + per-node & 512 & False & 0 & N/A & N/A & 0.000664 & 0.9 & none & 0.2 \\
GraphSAGE + column-wise & 1024 & False & 0 & N/A & N/A & 0.000469 & 0.9 & none & 0.2 \\
GraphSAGE + LSTM & 1024 & False & 0 & N/A & N/A & 0.000962 & 0.9 & none & 0.2 \\
GraphSAGE + Transformer & 512 & True & 0 & 2 & N/A & 0.000768 & 1.0 & none & 0.2 \\
\midrule
GAT + per-node & 1024 & False & 0 & N/A & 2 & 0.000002 & 0.90 & none & 0.25 \\
GAT + column-wise & 1024 & False & 0 & N/A & 2 & 0.000004 & 0.95 & none & 0.2 \\ 
GAT + LSTM & 1024 & False & 0 & N/A & 2 & 0.000006 & 0.95 & none & 0.25 \\
GAT + Transformer & 1024 & False & 0 & 2 & 2 & 0.000001 & 1.00 & norm & 0.2 \\
\bottomrule
\end{tabular}
    \caption{Tuned hyperparameters used in the model ablation experiment on the fusion dataset. MSE loss is used for this dataset.}
    \label{tab:hp-fusion}
\end{table*}

\begin{table*}[t]
    \centering
    \footnotesize
    \caption{Similar to \cref{tab:fusion-and-window-per-app} but on the manual split.
    }
    \label{tab:fusion-and-window-per-app-manual-split}
    \begin{tabular}{@{}lrrrrrrrr@{}}
\toprule
{} & \multicolumn{4}{c}{\textbf{Tile-Size}}
   & \multicolumn{4}{c}{\textbf{Fusion}} \\
\cmidrule(r){2-5}
\cmidrule(l){6-9}
{} & \multicolumn{2}{l}{\textbf{Tile-Size APE}}
   & \multicolumn{2}{l}{\textbf{Kendall's $\tau$}}
   & \multicolumn{2}{l}{\textbf{MAPE}}
   & \multicolumn{2}{l}{\textbf{Kendall's $\tau$}} \\
\cmidrule(r){2-3}
\cmidrule(lr){4-5}
\cmidrule(lr){6-7}
\cmidrule(l){8-9}
{} & \textbf{Learned}
   & \textbf{Analytical}
   & \textbf{Learned}
   & \textbf{Analytical}
   & \textbf{Learned}
   & \textbf{Analytical}
   & \textbf{Learned}
   & \textbf{Analytical} \\
\midrule
\textbf{Ranking}   &     9.5 &   1.4 &  0.81 &   0.71   &   10.8 &    10.7 &   0.72 &   0.81 \\
\textbf{Feats2Wave}   & 16.9 &   1.2 &  0.71 &   0.83   &    9.6 &    72.4 &   0.59 &   0.72 \\
\textbf{ImageEmbed}   &  5.7 &   5.6 &  0.81 &   0.75   &   11.4 &    14.6 &   0.90 &   0.90 \\
\textbf{SmartCompose}   & 3.2 &  1.6 &  0.67 &   0.76   &    6.6 &    40.2 &   0.96 &   0.95 \\
\textbf{WaveRNN 1}   &   7.0 &   2.6 &  0.66 &   0.81   &    2.7 &     8.8 &   0.97 &   0.95 \\
\textbf{WaveRNN 2}   &   3.4 &   4.4 &  0.72 &   0.68   &    2.8 &    10.3 &   0.97 &   0.94 \\
\midrule

\textbf{Median}     &   6.3 &   2.1 &  0.71 &   0.75 &    8.1 &    12.6 &     0.93 &   0.92   \\
\textbf{Mean}       &   6.4 &   2.3 &  0.73 &   0.75 &    6.2 &    18.1 &     0.84 &   0.88   \\
\bottomrule
\end{tabular}

\end{table*}

\end{document}